\documentclass[pra,aps,reprint,twocolumn,nosuperscriptaddress,10pt,nofootinbib,longbibliography,floatfix]{revtex4-2}

\usepackage[utf8]{inputenc}
\usepackage[T1]{fontenc}
\usepackage[english]{babel}
\usepackage{graphicx,epsfig}
\graphicspath{{images/}}
\usepackage{color}
\usepackage[usenames,dvipsnames]{xcolor}
\usepackage{rotating}

\usepackage{longtable}

\usepackage{physics}
\usepackage{amsmath,amssymb,amsthm}
\usepackage{bbm,dsfont} 
\usepackage{stmaryrd}
\usepackage{enumitem}
\usepackage[caption=false]{subfig}
\usepackage[normalem]{ulem}

\usepackage{mathtools}
\usepackage{cancel}

\definecolor{linkcolor}{rgb}{0,0,0.6}	
\usepackage[colorlinks=true,pdfstartview=FitV,linkcolor=linkcolor,citecolor=linkcolor,urlcolor=linkcolor,hyperindex=true,hyperfigures=false]{hyperref}

\usepackage{numprint}

\usepackage[bb=boondox]{mathalfa}

\usepackage{upgreek}

\usepackage[left=15mm,right=15mm,top=30mm, columnsep=15pt]{geometry}
\usepackage{adjustbox}
\usepackage{relsize}
\usepackage{placeins}
\usepackage{hyperref}
\usepackage{extarrows}
\usepackage{csquotes}

\usepackage{qcircuit}

\newcommand*{\e}{\text{e}}

\newcommand{\kket}[1]{{|#1 \rangle \!\rangle}}
\newcommand{\bbra}[1]{{\langle \!\langle #1 |}}
\renewcommand{\ket}[1]{{|#1 \rangle}}

\renewcommand{\ketbra}[2]{{|#1 \rangle \!\langle #2|}}
\def\kketbra#1#2{\mathinner{|{#1}\rangle\!\rangle\!\langle\!\langle{#2}|}}

\newcommand{\id}{\mathbbm{1}}

\newcommand{\HS}{\mathcal{H}}
\renewcommand\L{\mathcal{L}}

\renewcommand\L{\mathcal{L}}

\makeatletter 
    
\renewcommand\onecolumngrid{
\do@columngrid{one}{\@ne}%
\def\set@footnotewidth{\onecolumngrid}
\def\footnoterule{\kern-6pt\hrule width 1.5in\kern6pt}%
}

\renewcommand\twocolumngrid{
        \def\footnoterule{
        \dimen@\skip\footins\divide\dimen@\thr@@
        \kern-\dimen@\hrule width.5in\kern\dimen@}
        \do@columngrid{mlt}{\tw@}
}%

\makeatother    

\begin{document}

\title{
Witnessing the architecture of quantum circuits}

\author{Raphaël Mothe}
\affiliation{Naturwissenschaftlich-Technische Fakultät, Universität Siegen, Walter-Flex-Stra\ss e 3, 57068 Siegen, Germany}

\author{Otfried Gühne}
\affiliation{Naturwissenschaftlich-Technische Fakultät, Universität Siegen, Walter-Flex-Stra\ss e 3, 57068 Siegen, Germany}

\date{\today}


\begin{abstract}
Determining whether a target unitary can be implemented within a prescribed quantum circuit architecture is a fundamental problem in quantum information, with direct implications for optimisation and compilation of quantum circuits, and hardware-efficient quantum computation. While existing synthesis and compilation methods are primarily constructive, they generally do not provide rigorous certificates that a unitary cannot be realised using given implementation resources. Here we introduce a general framework to define quantum circuit architecture witnesses, which certify the incompatibility of a unitary transformation with a specified quantum circuit architecture. We formulate the witness construction as a semidefinite program by maximising the fidelity between the Choi state of the target unitary and those of tested circuits. The resulting witnesses provide practical and quantitative certificates of incompatibility, implying lower bounds on implementation resources such as the gate count or circuit depth, and can also be used experimentally to benchmark quantum devices by certifying that an implemented unitary channel goes beyond the capabilities of a given circuit architecture. For Clifford unitaries, we exploit the stabiliser formalism to reduce the construction to linear programming, enabling both more efficient numerical certification for circuits containing on the order of seven two-qubit gates, and analytical witnesses for some families of architectures made of an arbitrary number of gates.

\end{abstract}


\maketitle

{\it Introduction.---}
Quantum circuits constitute the standard model for quantum computation and quantum information processing, providing a structured sequence of quantum gates that transform input quantum systems into outputs \cite{deutsch85,Barenco95}. Beyond their role as a computational model, quantum circuits offer a natural framework for quantifying the resources required to implement a quantum operation. Indeed, while any unitary transformation can in principle be decomposed into elementary gates, different unitaries may require vastly different resources depending on the available architecture. Examples of resources include restrictions to nearest-neighbour interactions, limitations on the number of two-qubit gates, constraints on circuit depth, or the availability of experimentally expensive operations \cite{bernstein93,Chuang98,cleve00,nielsenchuang01,fenner03}. Such restrictions are of central importance both theoretically, where they determine the computational power of a model, and experimentally, where they directly affect the feasibility and fidelity of implementations. For instance, allowing interactions beyond nearest neighbours can qualitatively increase the computational power of a gate set, as illustrated by the role of non-nearest-neighbour interactions in the universality of matchgate circuits \cite{valiant01,valiant02,jozsa08}.

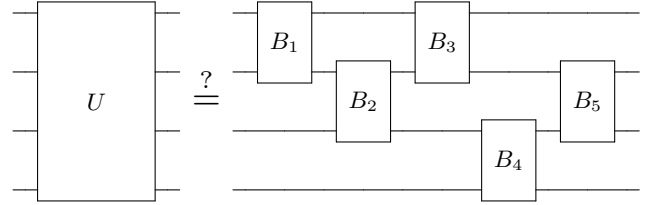
\begin{figure}[t!]
\begin{flushright}
\centering
$\Qcircuit @C=0.5em @R=1.5em {
  & \qw & \ghost{\rule{1.5em}{0pt}U\rule{1.5em}{0pt}} &\qw &\qw &          &&\qw & \ghost{B_1}        &\qw& \qw               & \qw & \ghost{B_3}           &     \qw           &\qw&\qw &\qw &\qw&\\
  & \qw & \ghost{\rule{1.5em}{0pt}U\rule{1.5em}{0pt}} &\qw & \qw & \push{\smash{\raisebox{-1.5em}{\scalebox{1.5}{$\overset{?}{=}$}}}} &&\qw & \multigate{-1}{B_1}&\qw& \ghost{B_2}         & \qw & \multigate{-1}{B_3} &     \qw           &\qw& \ghost{B_5} &\qw &\qw&\\
  & \qw & \ghost{\rule{1.5em}{0pt}U\rule{1.5em}{0pt}} &\qw  & \qw &       &&\qw & \qw              &\qw& \multigate{-1}{B_2} & \qw & \qw                   &\ghost{B_4}        &\qw& \multigate{-1}{B_5}&\qw &\qw&\\ 
  & \qw & \multigate{-3}{\rule{1.5em}{0pt}U\rule{1.5em}{0pt}} &\qw  & \qw&&& \qw & \qw              &\qw& \qw               & \qw & \qw                     &\multigate{-1}{B_4}&\qw&\qw &\qw &\qw& 
}$
\end{flushright}
\normalsize
    \caption{Illustration of a potential decomposition of the unitary transformation $U$ acting on four qubits, into a quantum circuit made of 5 two-qubit gates $B_1,\ldots,B_5$ composed sequentially.}
    \label{fig:causal_dec}
\end{figure}

Understanding which quantum operations are easy or difficult to implement under given architectural constraints is therefore a fundamental problem in quantum information science. From a theoretical perspective, this question is closely related to establishing lower bounds on resources such as gate count or circuit depth \cite{cleve00,nielsenchuang01,fenner03}. From an experimental perspective, shallower circuits and circuits involving fewer entangling gates are generally less susceptible to noise and decoherence, making resource-efficient implementations highly desirable \cite{Magesan12}. This motivates the following question: how to certify that a given target unitary is impossible to implement within a specified quantum circuit architecture? While many techniques have been developed for quantum circuit synthesis \cite{yan25}, or quantum circuit compilation and optimisation \cite{chong17,maronese22,karuppasamy25}, these methods are generally designed to construct implementations rather than certify their impossibility. In this work, we address this question by studying the problem of certifying the incompatibility of a unitary transformation with a given quantum circuit architecture, see an illustration in Fig.~\ref{fig:causal_dec}.

We develop a framework for constructing circuit architecture witnesses, which certify the incompatibility of a unitary channel with a given circuit architecture, analogously to how entanglement witnesses \cite{Horodecki96,Terhal00} certify entanglement of quantum states and causal witnesses \cite{Araujo15, branciard16} certify causal non-separability of process matrices. By evaluating fidelities between the Choi state of the target unitary and those of candidate circuits compatible with a given architecture, our framework reduces the problem of certifying circuit compatibility to a separability problem. This reduction is of independent interest, as a wide range of results and approximation techniques are available for separability problems, and related problems, such as the quantum marginal problem \cite{yu21}, can likewise be formulated in terms of separability. At the same time, the framework provides a practical and quantitative criterion for certifying architectural incompatibility. We derive two key consequences from these witnesses. First, they can be used to prove theoretically that certain unitaries cannot even be approximated beyond a given fidelity by circuits with a prescribed architecture, thereby yielding lower bounds on implementation resources such as circuit depth. Second, they provide an experimentally accessible benchmarking tool: an experimenter can use the witness to certify that an implemented transformation goes beyond what can be achieved by a specific quantum circuit architecture, demonstrating that the transformation requires a more complex or structurally refined architecture.

\vspace*{5pt}

\begin{figure}[t]
\centering
\includegraphics[width=\columnwidth]{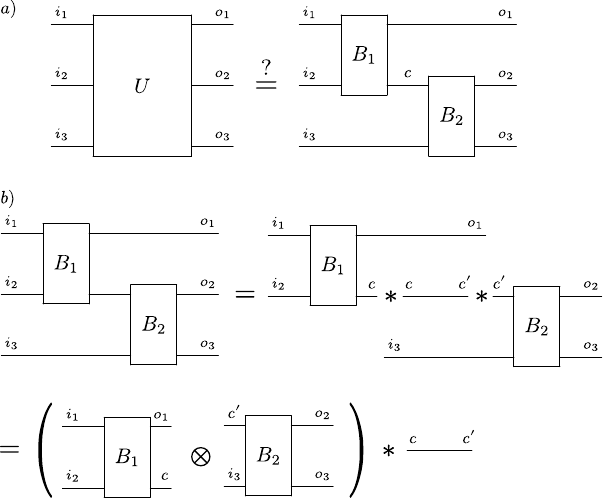}
    \caption{a) Minimal example of the architecture incompatibility problem, between a tripartite unitary $U$ and the quantum circuit architecture $\Gamma=(\{1,2\},\{2,3\})$. The input and output systems are assumed to be all qubits. Furthermore, when defining a circuit architecture, we assume that all internal wires, i.e.~here the wire labelled $c$ which connects $B_1$ and $B_2$, is also carrying a qubit. This assumption is not restricting the set of allowed unitary gates for each $B_k$. b) Illustration of Eq.~\eqref{eq:rewrite_objective_function}, which consists in rewriting in the Choi picture the composition of gates $B_1$ and $B_2$ as a tensor product of the Choi matrices of the two gates.}
    \label{fig:U_2_gates}
\end{figure}

\textit{Architecture incompatibility witness framework.---}\label{sec:witness_def} For the sake of simplicity we will consider a minimal example of the architecture incompatibility problem which we introduced in Fig.~\ref{fig:causal_dec}. We will then explain how the framework can be extended to the general case of any quantum circuit architecture\footnote{We note that the nomenclature ``quantum circuit architecture'' was first mentioned in the context of higher-order quantum circuits, also known as quantum combs \cite{Chiribella08}, while in this work it refers to the architecture of standard quantum circuits.}. Consider a target unitary operator $U:\HS^{i_1 i_2 i_3}\to\HS^{o_1 o_2 o_3}$ that transforms three input systems respectively attached to input Hilbert spaces $\HS^{i_k}$ into three output systems respectively attached to output Hilbert spaces $\HS^{o_k}$, for $k=1,2,3$, and with the short-hand notation $\HS^{X_1X_2 X_3}:=\HS^{X_1}\otimes\HS^{X_2}\otimes \HS^{X_3}$. For the sake of simplicity we assume that all input and output systems are qubits.
One would like to determine whether this target unitary $U$ can be decomposed as a sequential composition of 2 two-qubit unitary gates $B_1$ and $B_2$, where $B_1$ is applied on the first two qubit systems before $B_2$ is applied on the last two qubit systems, see an illustration in Fig.~\ref{fig:U_2_gates}. Importantly, we assume that the internal wire, which connects the two unitary gates, is carrying a qubit system. This assumption turns out to be crucial to forbid trivial realisations of the target unitary as we explain in Appendix~\ref{app:fixing_dim_internal_wire}. The circuit architecture can then described by the list $\Gamma=(\{1,2\},\{2,3\})$, where the first element $\{1,2\}$ denotes the labels of the systems (here the first two systems) on which is applied the gate $B_1:\HS^{B_1^I}\to\HS^{B_1^O}$, and the second element $\{2,3\}$ denotes the labels of the systems (here the last two systems) on which is applied the gate $B_2:\HS^{B_2^I}\to\HS^{B_2^O}$. If such a decomposition exist, we say that the target unitary $U$ is compatible with the circuit architecture $\Gamma$.

Throughout this work we will describe quantum circuits made of the composition of different quantum gates. A handy tool to describe such objects is the so-called Choi–Jamio{\l}kowski isomorphism. It allows us to describe any unitary channel $B:\HS^X\to\HS^Y$ by its Choi vector as the ``double-ket'' vector $\kket{B} \coloneqq \sum_i \ket{i}^X\otimes B\ket{i}^X \ \in \HS^{XY}$, with $\{\ket{i}^X\}_i$ denoting the computational basis of $\HS^X$. In the Choi-Jamio\l{}kowski representation, a completely positive and trace-preserving (CPTP) map, also known as a quantum channel, $\mathcal{C}:\L(\mathcal{H}^X)\to\L(\mathcal{H}^Y)$ is described by its Choi operator $C\coloneqq\sum_{ij}\ketbra{i}{j}^{X}\otimes\mathcal{C}(\ketbra{i}{j}^{X}) \in\L(\mathcal{H}^{XY})$, and satisfies $C\geq0$ (complete positivity) and $\Tr_Y[C]=\id_X$ (trace preservation), where $\id_X$ is the identity operator acting on $\mathcal{H}^X$. Importantly, the transformation can be reversed for any unitary or quantum channel, so that the unitary or quantum channel is equivalently described by its Choi vector or operator respectively.

Let $\kket{U} \in \HS^{i_1 i_2 i_3 o_1 o_2 o_3}$ be the Choi vector of the target unitary $U$ acting on 3 qubits, with norm $8$. Our approach to solve the architecture incompatibility problem is to optimise the fidelity between $\kket{U}$ and the Choi vector $\kket{B_1}*\kket{B_2}$ of the quantum circuit defined by the composition of $B_1$ and $B_2$ according to $\Gamma$, where $*$ denotes the so-called link product~\cite{Chiribella08,Chiribella09,wechs21} which expresses the composition of maps in the Choi picture. Denoting $\HS^c$ the common Hilbert space between $B_1$ and $B_2$, it is defined as $\kket{B_1} * \kket{B_2} \coloneqq  \big( \id \otimes \bbra{\id}^{cc} \big) \big(\kket{B_1} \otimes \kket{B_2}\big)$. Solving the architecture incompatibility problem thus boils down to solving the following fidelity optimisation problem:
    \begin{align}
         \textbf{given}\ \ & U, \Gamma \nonumber \\
    \textbf{max}_{B_1,B_2}\ \   & \frac{1}{64}\Tr\left[\kketbra{U}{U} \cdot \kketbra{B_1}{B_1}*\kketbra{B_2}{B_2} \right] \nonumber \\
    \textbf{s.t.}\ \  & B_1,B_2 \ \text{are unitary operations applied}\nonumber\\
    &\text{according to the circuit architecture $\Gamma$.}
    \label{eq:max_overlap_gen}
    \end{align}
We denote by $\alpha_\Gamma(U)$ the optimised fidelity in Eq.~\eqref{eq:max_overlap_gen}, it is normalised and thus ranges between 0 and 1. If $\alpha_\Gamma(U)<1$, then the target unitary is incompatible with the circuit architecture $\Gamma$, as no choice of unitary gates $B_1,B_2$ arranged according to $\Gamma$ and under the assumptions stated above can realise the target unitary $U$. If $\alpha_\Gamma(U)=1$, then the target unitary $U$ is compatible with the circuit architecture $\Gamma$ and one can extract the corresponding compatible quantum circuit from the optimisation.

In analogy to entanglement witnesses \cite{Horodecki96,Terhal00}, when $\alpha_\Gamma(U)<1$, one can build a witness $\mathcal{W} \in \L(\HS^{i_1i_2i_3 o_1o_2 o_3})$ from the optimised fidelity as 
\begin{align}
    \mathcal{W} = \alpha_\Gamma(U) \id - \frac{1}{8}\kketbra{U}{U}.
    \label{eq:def_witness}
\end{align}
By construction $\Tr(\mathcal{W}\kketbra{V}{V})\geq 0$ for any normalised Choi state $\kketbra{V}{V}$ representing a unitary $V$ compatible with $\Gamma$, while $\Tr(\mathcal{W}\kketbra{V}{V}) < 0$ witnesses that the unitary $V$ is incompatible with the circuit architecture $\Gamma$. The method has experimental implications: measuring experimentally the fidelity between a target unitary $U$ and an experimental realisation of it \cite{pallister17,bavaresco18,friis19} that is higher than $\alpha_\Gamma(U)$ certifies that the experimental realisation is incompatible with the quantum circuit architecture $\Gamma$. Furthermore, the method is inherently robust to experimental noise, as the latter typically lowers the measured fidelity, but for realistic noise levels the fidelity is expected to remain above the certification threshold.

Note that solving Eq.~\eqref{eq:max_overlap_gen} is in general hard as the optimisation problem is polynomial in the Choi matrices of the gates $B_k$, and as imposing that the gates $B_k$ are unitaries is a non-convex constraint. In what follows we are thus going to build witnesses for various target unitaries $U$ and circuit architectures $\Gamma$ by first connecting it to a separability problem, and then relaxing the problem using known
entanglement criteria, and thus upper bounding $\alpha_\Gamma(U)$. Indeed,  as we will see, replacing $\alpha_\Gamma(U)$ by a strictly lower than 1 upper bound in Eq.~\eqref{eq:def_witness} is enough to construct a strong witness.

\textit{SDP relaxation for general target unitaries.---}We start by rewriting the optimisation problem of Eq.~\eqref{eq:max_overlap_gen}, before relaxing it to a SDP problem. To preserve the assumption that the internal wire in the circuit corresponds to a qubit system, we reformulate the optimisation problem to keep explicitly track of this wire. This is done by ``opening'' the internal wire and imposing that it corresponds to an identity wire carrying a qubit, which is described in the Choi picture by $\kket{\id}^{cc'}$, linking the output Hilbert space $\HS^c$ of $B_1$ to the input Hilbert space $\HS^{c'}$ of $B_2$ (see Fig.~\ref{fig:U_2_gates}b for an illustration). The resulting objective function can thus be written as 
\begin{align}
    &\Tr\left[\kketbra{U}{U} \cdot \kketbra{B_1}{B_1}*\kketbra{B_2}{B_2} \right] = \notag \\ & \quad \Tr\left[(\kketbra{U}{U}\otimes \kketbra{\id}{\id}^{cc'})\cdot (\kketbra{B_1}{B_1}\!\otimes\!\kketbra{B_2}{B_2}) \right], \label{eq:rewrite_objective_function}
\end{align}
see Appendix~\ref{app:rewriting_relaxation_SDP} for more details. Despite this reformulation, the optimisation problem remains non-convex, both because of the bilinear dependence on the gate variables $B_1$ and $B_2$ and the unitary constraints. We thus relax the problem to a semidefinite program by introducing a positive semidefinite variable $\tau \in \L(\HS^{B_1^{IO}B_2^{IO}})$ subject to positive partial transpose (PPT) condition across the bipartition $B_1^{IO}|B_2^{IO}$, and CPTP constraints on its marginals across this bipartition. As shown in Appendix~\ref{app:rewriting_relaxation_SDP}, this defines the following SDP problem
    \begin{align}
         \textbf{given}\ \ & U, \Gamma \nonumber \\
    \textbf{max}_\tau\ \   & \frac{1}{64}\Tr\left[(\kketbra{U}{U}\otimes \kketbra{\id}{\id})\cdot \tau\right] \nonumber \\
    \textbf{s.t.}\ \  & \tau \geq 0, \  \Tr[\tau] = 16, \nonumber \\
    &\tau^{T_{{B}^{IO}_1}}\geq0,\nonumber \\
    & \forall \, k=1,2, \ \Tr_{B_k^O}[\tau] = \Tr_{B_k^{IO}}[\tau]\otimes \frac{\id_{B_k^I}}{4}.
    \label{eq:max_overlap_cut_SDP}
    \end{align}
    We show in Appendix~\ref{app:rewriting_relaxation_SDP_general} that the technique can be extended to general circuit architectures made of $n$ systems and $m$ gates, and that the above approach also allows to phrase the optimisation problem as a SDP. In Appendix~\ref{app:alternative_relaxation}, we propose an alternative SDP relaxation of Eq.~\eqref{eq:max_overlap_gen}, which builds on the quantum comb formalism \cite{Chiribella08,Chiribella09}. The size of the variable $\tau$ scales exponentially with the number of gates of the architecture, which makes the optimisation on most architectures made of 4 two-qubit unitaries or more intractable with standard resources as we will see with practical examples. To partially address the scalability problem, we now turn to the particular case of target Clifford unitaries, and show that the SDP relaxations simplify into linear programming (LP) relaxations, which can be optimised over a higher number of gates.

\textit{LP relaxation for target Clifford unitaries.---}\label{sec:clifford_gates}We now focus on the quantum circuit architecture incompatibility of target Clifford unitaries, i.e.~unitaries that map Pauli operators to Pauli operators under conjugation \cite{gottesman97}. This simplifying assumption finds motivation in the fact that Clifford gates are ubiquitous in quantum information and quantum computing, in particular in the field of quantum error correction \cite{Roffe19}. Consider a Clifford target unitary operator
$U$. As recalled in Appendix~\ref{app:graph_stabiliser}, its Choi vector $\kket{U}$ is a non-normalised stabiliser state. Since $\kket{\id}$ is a non-normalised Bell state which is also a stabiliser state, the objective function in Eq.~\eqref{eq:max_overlap_cut_SDP} is the trace of the product between a (non-normalised) stabiliser state $\kketbra{U}{U}\otimes \kketbra{\id}{\id}$ and a positive semidefinite variable $\tau$. As any stabiliser state, the former is local-Clifford equivalent to a graph state $\ket{G}$ \cite{hein06}. Using the fact that any state can, by local operations, be depolarised
to a graph-diagonal state \cite{Jungnitsch11}, we show that without loss of generality $\tau$ can be taken to be diagonal in the graph-state basis $\{\ket{G_i}\}_i$ defined by $\ket{G}$, see Appendix~\ref{app:graph_stabiliser}. The optimisation problem of Eq.~\eqref{eq:max_overlap_cut_SDP} thus reduces to optimising over the diagonal coefficients of $\tau$ in this basis. Moreover, we show in Appendix~\ref{app:rewriting_SDP_to_LP} that all the linear constraints on $\tau$ defined in Eq.~\eqref{eq:max_overlap_cut_SDP} translates into linear constraints on the diagonal coefficients of $\tau$, thus the optimisation problem simplifies into a linear program (LP) which is formally presented in Eq.~\eqref{eq:max_overlap_cut_LP} of Appendix~\ref{app:rewriting_SDP_to_LP}. Before turning to concrete examples using the two numerical methods to compute non-trivial witnesses, we finally propose an analytical method to tackle the quantum circuit architecture incompatibility problem.

\textit{Analytical method for target Clifford unitaries.---}As already observed in Eq.~\eqref{eq:rewrite_objective_function}, the optimisation problem can be rewritten as a maximisation of the fidelity between the entangled state $\kketbra{U}{U}\otimes \kketbra{\id}{\id}$ and a separable state $\kketbra{B_1}{B_1}\otimes\kketbra{B_2}{B_2}$ across the bipartition $B_1^{IO}|B_2^{IO}$, with the extra constraints that $B_1$ and $B_2$ are unitaries. In Appendix~\ref{app:analytical_witness}, we show that for a Clifford unitary $U$ and for so-called staircase circuit architectures with no-signalling relations between input and output systems, one can get rid of the conditions imposing that $B_k$ are unitaries and derive non-trivial analytical upper bounds on Eq.~\eqref{eq:max_overlap_gen}. Interestingly, such upper bounds can be derived by counting the number of Bell pairs (defined across the bipartition $B_1^{IO}|B_2^{IO}$) that can be extracted from the stabiliser state using local (in the sense of the bipartition) unitaries \cite{Fattal04}. As we show in Appendix~\ref{app:analytical_witness_2}, such a technique applies for instance to the circuit architecture presented in Fig.~\ref{fig:U_2_gates}. In Appendix~\ref{app:analytical_witness_N}, we also present an example illustrating how this technique can be used to establish a non-trivial upper bound on fidelity in the case of a circuit architecture comprising an arbitrary number of gates, highlighting the merit of the analytical approach to circumvent the ``curse of dimensionality'' which is a bottleneck of the two numerical approaches presented above (LP and SDP). An interesting follow-up question is whether the technique can be generalised to any quantum circuit architecture, without assuming any no-signalling relation between the input and output systems.

\begin{figure}[t!]
\begin{flushright}
\centering
$\Qcircuit @C=1em @R=2em {
  & \qw_{i_1} & \qw               &\qw& \ghost{A_2}   & \qw & \qw_{o_1} && & \qw_{i_1} & \ghost{B_1}        &\qw& \qw               & \qw & \qw_{o_1} &\\
  &  \qw_{i_2} & \ghost{A_1}         &\qw& \multigate{-1}{A_2}         & \qw & \qw_{o_2} &\push{\smash{\raisebox{-0.35em}{\scalebox{1.5}{$\overset{?}{=}$}}}}& & \qw_{i_2} & \multigate{-1}{B_1}&\qw& \ghost{B_2}         & \qw & \qw_{o_2} &\\
  & \qw_{i_3} &  \multigate{-1}{A_1}&\qw& \qw& \qw & \qw_{o_3} && & \qw_{i_3} & \qw              &\qw& \multigate{-1}{B_2} & \qw & \qw_{o_3} &
}$
\end{flushright}
\normalsize
    \caption{Illustration of the incompatibility problem between the circuit architectures $\Gamma_{\text{target}}=(\{2,3\},\{1,2\})$ and $\Gamma_{\text{test}}=(\{1,2\},\{2,3\})$. See text for further details.}
    \label{fig:cascade_example_2_gates}
\end{figure}
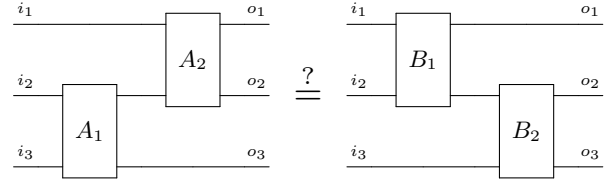

\textit{Examples I: Circuits made of two
unitaries.---}\label{sec:examples_cascade}For simplicity we start by considering the minimal non-trivial scenario with examples of target unitary transformations $U$ defined as the composition of 2 two-qubit unitary gates $A_1$ and $A_2$ applied on three qubits and compatible with the circuit architecture $\Gamma_{\text{target}}=(\{2,3\},\{1,2\})$, i.e., $A_1$ is first applied on the last two qubits before $A_2$ is applied on the first two qubits. We study the compatibility of $U$ with the circuit architecture $\Gamma_{\text{test}}=(\{1,2\},\{2,3\})$, made of 2 gates $B_1$ and $B_2$, see Fig.~\ref{fig:cascade_example_2_gates} for an illustration of the incompatibility problem.

The SDP defined in Eq.~\eqref{eq:max_overlap_cut_SDP} for general target unitaries, as well as the LP defined in Eq.~\eqref{eq:max_overlap_cut_LP} for target Clifford unitaries, can provide a non-trivial upper bound of the optimised fidelity $\alpha_{\Gamma_{\text{test}}}(U)$. We report in Table~\ref{tab:cascade_example_2_gates} the upper bounds obtained by solving the LP optimisation for various choices of target Clifford unitary (i.e., for Clifford gates $A_1$ and $A_2$), or by solving the SDP optimisation for target non-Clifford unitaries. In all the cases considered the upper bound is strictly lower than 1, thus the approach certifies the incompatibility of the corresponding $U$ with the circuit architecture $\Gamma_{\text{test}}$. 

\begin{table}[t]
\begin{tabular}{|c|c|c|}
\hline
$A_1$ & $A_2$ & $\alpha_{\Gamma_{\text{test}}}(U) $ \\
\hline & & \\[-8pt]
SWAP & SWAP & 0.25 \\
SWAP & CNOT & 0.5 \\
CNOT & CNOT & 0.5 \\
$\e^{i\frac{\pi}{4}XX}$ & $\e^{i\frac{\pi}{4}YY}$ & 0.5 \\
$\e^{i\frac{9\pi}{32}XX}$ & $\e^{i\frac{9\pi}{32}YY}$ & $\simeq 0.6366$
\\
\hline
\end{tabular}
\caption{Upper bound on the optimised fidelity obtained by solving numerically the LP defined in Eq.~\eqref{eq:max_overlap_cut_LP} for target Clifford unitaries (four first lines of the table) or alternatively Eq.~\eqref{eq:max_overlap_cut_SDP} for target non-Clifford unitaries (last line of the table). The considered target unitaries $U$ are made of two gates $A_1$ and $A_2$ composed according to the circuit architecture $\Gamma_{\text{target}}=(\{2,3\},\{1,2\})$ and a tested quantum circuit made of 2 gates compatible with $\Gamma_{\text{test}}=(\{1,2\},\{2,3\})$. As it can be found by running a see-saw algorithm, all the upper bounds found here are tight (up to a numerical precision of $10^{-5}$).}
\label{tab:cascade_example_2_gates}
\end{table}

That these examples of $U$ are incompatible with $\Gamma_{\text{test}}$ can be easily understood as by construction the circuit architecture $\Gamma_{\text{target}}$ allows for signalling from $\HS^{i_3}$ to $\HS^{o_1}$ while this is prevented in $\Gamma_{\text{test}}$. Still the approach provides a quantitative study of the incompatibility of a target unitary with a circuit architecture, which is not possible by simply checking the no-signalling relations between input and output systems of $U$.

Finally, we consider the example where the target transformation $U$ is a Toffoli (or CCNOT) gate and obtain solving Eq.~\eqref{eq:max_overlap_cut_SDP} an upper bound of $\simeq 0.7285$ on $\alpha_{\Gamma_{\text{test}}}(U)$, which shows that the Toffoli gate is not compatible with the circuit architecture $\Gamma_{\text{test}}=(\{1,2\},\{2,3\})$. We also verify that the bound is actually tight (up to $10^{-5}$) with a see-saw optimisation. This is in agreement with Ref.~\cite{yu13} where it is shown that 5 two-qubit unitaries are required to implement a Toffoli gate. This example highlights that the framework introduced here allows us to provide lower bounds on the circuit depth of given unitary transformations. Indeed, we obtained for all the six possible circuit architectures made of 2 two-qubit gates the same upper bound on the optimised fidelity, thus we can rule out the possibility that the Toffoli is realisable with 2 two-qubit gates applied on 3 qubits. We conclude that its circuit depth when the allowed resources are two-qubit unitaries is at least 3.

\begin{figure}[t!]
\begin{flushright}
\centering
$\Qcircuit @C=0.75em @R=2em {
  & \qw_{i_1} & \ghost{A_1}        &\qw& \qw               &\qw & \ghost{A_3} & \qw &  \qw_{o_1} && & \qw_{i_1} & \qw               &\qw& \ghost{B_2}   & \qw& \qw&\qw&\qw_{o_1} &\\
  & \qw_{i_2} & \multigate{-1}{A_1}&\qw& \ghost{A_2}         & \qw & \multigate{-1}{A_3}&\qw &\qw_{o_2}  &\push{\smash{\raisebox{-0.35em}{\scalebox{1.5}{$\overset{?}{=}$}}}}& & \qw_{i_2} & \ghost{B_1}         &\qw& \multigate{-1}{B_2}         & \qw & \ghost{B_3}         &\qw& \qw_{o_2} &\\
  & \qw_{i_3} & \qw              &\qw& \multigate{-1}{A_2} & \qw & \qw & \qw & \qw_{o_3} && & \qw_{i_3} &  \multigate{-1}{B_1}&\qw& \qw& \qw & \multigate{-1}{B_3}&\qw& \qw_{o_3} &
}$
\end{flushright}
\normalsize
    \caption{Illustration of the incompatibility problem between the circuit architectures $\Gamma_{\text{target}} = (\{1,2\},\{2,3\},\{1,2\})$ and $\Gamma_{\text{test}} = (\{2,3\},\{1,2\},\{2,3\})$. We note that when exchanging the qubits
    2 and 3, then the circuit on the left-hand side has two gates beyond nearest neighbours interactions, while the circuit on the right-hand side has only one, which highlights the physical difference between the two corresponding quantum circuit architectures.}
    \label{fig:example_3_gates}
\end{figure}
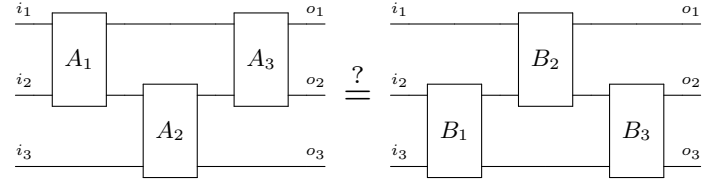

\textit{Examples II: Circuits made of three unitaries.---}\label{sec:examples_general}We now move on to target unitaries $U$ applied on three qubits made of three two-qubit unitary gates $A_1, A_2$ and $A_3$ composed according to the circuit architecture $\Gamma_{\text{target}} = (\{1,2\},\{2,3\},\{1,2\})$ and study the incompatibility of the resulting unitary quantum circuits with the counterpart circuit architecture $\Gamma_{\text{test}} = (\{2,3\},\{1,2\},\{2,3\})$, made of three gates $B_1$, $B_2$ and $B_3$, see Fig.~\ref{fig:example_3_gates}. We highlight that the no-signalling argument that could be used above to certify incompatibility is no longer available as the circuit architectures $\Gamma_{\text{target}}$ and $\Gamma_{\text{test}}$ allows for signalling from all input to all output systems.

We present in Table~\ref{tab:example_3_gates} two examples of unitaries $A_1,A_2$ and $A_3$ whose composition according to $\Gamma_{\text{target}}=(\{1,2\},\{2,3\},\{1,2\})$ defines a unitary $U$ incompatible with $\Gamma_{\text{test}}=(\{1,2\},\{2,3\},\{1,2\})$.

\begin{table}[t]
\begin{tabular}{|c|c|c|c|}
\hline
$A_1$ & $A_2$ & $A_3$ & $\alpha_{\Gamma_{\text{test}}}(U)$ \\
\hline & & & \\[-8pt]
$\e^{i\frac{\pi}{4}XX}$ & $\e^{i\frac{\pi}{4}YY}$ & $\e^{i\frac{\pi}{4}ZZ}$ & $0.5$ \\
$\e^{i\frac{9\pi}{32}XX}$ & $\e^{i\frac{9\pi}{32}YY}$ & $\e^{i\frac{9\pi}{32} ZZ}$ & $ \leq 0.6798$
\\
\hline
\end{tabular}
\caption{Upper bound on the optimised fidelity obtained by solving numerically Eq.~\eqref{eq:max_overlap_cut_LP} for a target Clifford unitary (first line of the table) or alternatively Eq.~\eqref{eq:max_overlap_cut_SDP} for target non-Clifford unitaries (second line of the table). The considered target unitaries are made of three two-qubit unitary gates $A_1,A_2$ and $A_3$ composed according to the circuit architecture $\Gamma_{\text{target}}=(\{1,2\},\{2,3\},\{1,2\})$ and the tested quantum circuit is made of 3 gates compatible with the architecture $\Gamma_{\text{test}}=(\{2,3\},\{1,2\},\{2,3\})$. Thanks to a see-saw algorithm, we show that the upper bound on the first line is tight (up to $10^{-6}$ numerical precision), while only $\simeq 0.6502$ is achieved for the case of the second line.}
\label{tab:example_3_gates}
\end{table}
Interestingly these two examples contrast with results previously obtained on so-called matchgate unitaries for which it was shown in \cite{Kokcu22,Camps22} that the architectures $\Gamma_{\text{target}}$ and $\Gamma_{\text{test}}$ are always compatible, see also Ref.~\cite{morralyepes25}.

We again considered the example of the Toffoli gate, already studied above for test circuit architectures made of two gates. Here for the test circuit architecture $\Gamma_{\text{test}}=(\{2,3\},\{1,2\},\{2,3\})$ we obtain an upper bound of $\simeq 0.857$ on $\alpha_{\Gamma_{\text{test}}}$ with SCS \cite{ocpb16} (with a numerical precision of $10^{-4}$). Checking analogously the incompatibility of the Toffoli gate with all the 12 possible circuit architectures made of 3 two-qubit unitaries on 3 qubit systems, we can thus show that the circuit depth of the Toffoli gate is at least 4, which is once more in agreement with Ref.~\cite{yu13}.

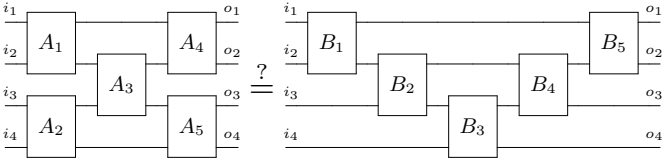
\begin{figure}[t!!!]
\resizebox{\columnwidth}{!}{
$\Qcircuit @C=0.5em @R=1em {
  &\qw_{i_1} & \ghost{A_1}        & \qw & \qw           &     \qw &\ghost{A_4}&\qw &\qw_{o_1}&& &\qw_{i_1} & \ghost{B_1}        &\qw& \qw               & \qw & \qw           &     \qw           &\qw&\qw &\ghost{B_5} &\qw &\qw_{o_1}&\\
  &\qw_{i_2} & \multigate{-1}{A_1}& \qw         & \ghost{A_3} &     \qw           & \multigate{-1}{A_4} &\qw &\qw_{o_2}&\push{\smash{\raisebox{-1.5em}{\scalebox{1.5}{$\overset{?}{=}$}}}} & &\qw_{i_2} & \multigate{-1}{B_1}&\qw& \ghost{B_2}         & \qw & \qw &     \qw           &\ghost{B_4} &\qw & \multigate{-1}{B_5} &\qw &\qw_{o_2}&\\
 &\qw_{i_3}     & \ghost{A_2} & \qw & \multigate{-1}{A_3}                           &\qw &\ghost{A_5} &\qw &\qw_{o_3}&& &\qw_{i_3} & \qw              &\qw& \multigate{-1}{B_2} & \qw & \ghost{B_3}                   &\qw        &\multigate{-1}{B_4} &\qw &  \qw &\qw &\qw_{o_3}&\\ 
  & \qw_{i_4}              & \multigate{-1}{A_2}               & \qw & \qw                     &\qw & \multigate{-1}{A_5}&\qw &\qw_{o_4}&& & \qw_{i_4} & \qw              &\qw& \qw               & \qw &\multigate{-1}{B_3}                     &\qw &\qw &\qw &\qw &\qw &\qw_{o_4}& 
}$}
    \caption{Illustration of the incompatibility problem between the circuit architectures $\Gamma_{\text{target}} = (\{1,2\},\{3,4\},\{2,3\},\{1,2\},\{3,4\})$ and $\Gamma_{\text{test}} = (\{1,2\},\{2,3\},\{3,4\},\{2,3\},\{1,2\})$.}
    \label{fig:example_5_gates}
\end{figure}

\textit{Examples III: Circuits made of five
unitaries.---}We consider the example of a Clifford target unitary $U$ applied on 4 qubits made of five two-qubit Clifford unitary gates $A_1 = \e^{i\frac{\pi}{4}XX}, A_2 = \e^{i\frac{\pi}{4}XX}, A_3 = \e^{i\frac{\pi}{4}YY}, A_4 = \e^{i\frac{\pi}{4}XX}$ and $A_5 = \e^{i\frac{\pi}{4}ZZ}$ composed according to the circuit architecture $\Gamma_{\text{target}} = (\{1,2\},\{3,4\},\{2,3\},\{1,2\},\{3,4\})$ and show the incompatibility of the resulting unitary quantum circuits with the counterpart circuit architecture $\Gamma_{\text{test}} = (\{1,2\},\{2,3\},\{3,4\},\{2,3\},\{1,2\})$ made of 5 gates, see Fig.~\ref{fig:example_5_gates}.

Using a LP introduced in Eq.~\eqref{eq:max_overlap_cut_LP_general} in Appendix \ref{app:rewriting_SDP_to_LP}, one can prove the incompatibility as we obtain an upper bound on the optimised fidelity of $0.5$; this bound being tight, as demonstrated by a see-saw approach.

\textit{Conclusion.---}
\label{sec:conclusion}
We have introduced a framework to quantitatively witness that a given unitary transformation cannot be realised within a prescribed quantum circuit architecture by leveraging fidelity-based witnesses defined in terms of Choi states. This framework has significant implications for both theory and experiments: on the theoretical side, it provides rigorous proofs of incompatibility, while on the experimental side it establishes a benchmarking framework for quantum devices by linking experimentally accessible certification to the computational resources required to implement quantum operations. We leave as an interesting open question whether general circuit architectures can be ruled out analytically when the target unitary is a Clifford operation. To facilitate the practical implementation of this framework in experiments, it is also important to investigate how the number of measurements required to estimate the fidelity can be minimised, thereby enabling efficient measurement schemes for constructing witnesses that rule out a given architecture.

\begin{acknowledgments}
\vspace{-2mm}
We thank Alastair Abbott, Cyril Branciard, Sophia Denker, Simon Milz, Ties Ohst, Pierre Pocreau, Christian Roos,  Leonardo S. V. Santos, Roope Uola and Augustin Vanrietvelde for helpful discussions and comments on this work.
This research was funded by the Deutsche Forschungsgemeinschaft (DFG, German Research Foundation, project number 563437167), the Sino-German Center for Research Promotion (Project M-0294), and the German Federal Ministry of Research, Technology and Space (Project QuKuK, Grant 
No.\ 16KIS1618K and Project BeRyQC, Grant No.\ 13N17292). R. Mothe acknowledges the support 
from the Alexander von Humboldt Foundation.
\end{acknowledgments}

\bibliography{biblio_witness_causal}

\clearpage
\onecolumngrid
\appendix

\section*{Appendices}

\section{SDP relaxation of the architecture incompatibility optimisation problem}
\label{app:SDP_relaxation_1}

\subsection{Fixing the dimension of the internal wires of the quantum circuit architecture}\label{app:fixing_dim_internal_wire}

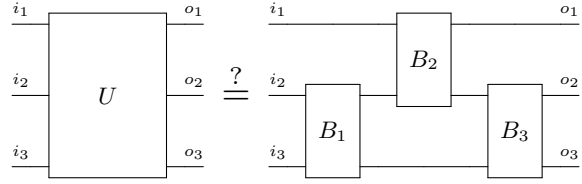
\begin{figure}[h!]
\begin{flushright}
\centering
$\Qcircuit @C=0.75em @R=2em {
  & \qw_{i_1} & \ghost{\rule{1.5em}{0pt}U\rule{1.5em}{0pt}} &\qw &\qw_{o_1} && & \qw_{i_1} & \qw               &\qw& \ghost{B_2}   & \qw& \qw&\qw&\qw_{o_1} &\\
  & \qw_{i_2} & \ghost{\rule{1.5em}{0pt}U\rule{1.5em}{0pt}} &\qw & \qw_{o_2}  &\push{\smash{\raisebox{-0.35em}{\scalebox{1.5}{$\overset{?}{=}$}}}}& & \qw_{i_2} & \ghost{B_1}         &\qw& \multigate{-1}{B_2}         & \qw & \ghost{B_3}         &\qw& \qw_{o_2} &\\
  & \qw_{i_3} & \multigate{-2}{\rule{1.5em}{0pt}U\rule{1.5em}{0pt}} &\qw  & \qw_{o_3} && & \qw_{i_3} &  \multigate{-1}{B_1}&\qw& \qw& \qw & \multigate{-1}{B_3}&\qw& \qw_{o_3} &
}$
\end{flushright}
\normalsize
    \caption{Illustration of the incompatibility problem between the tripartite unitary $U$ and $\Gamma_{\text{test}} = (\{2,3\},\{1,2\},\{2,3\})$.}
    \label{fig:example_3_gates_app}
\end{figure}

We start by explaining with a concrete example why it is crucial to fix the dimension of the internal wires of the quantum circuit architecture. We consider the example of the incompatibility problem between a tripartite unitary $U$ and the circuit architecture $\Gamma_{\text{test}} = (\{2,3\},\{1,2\},\{2,3\})$, see Fig.~\ref{fig:example_3_gates_app}.

As explained in the main text, the systems carried by all the wires, and in particular the internal ones connecting the gates $B_1,B_2$ and $B_3$ in Fig.~\ref{fig:example_3_gates_app}, are fixed to be qubits. Indeed, if the dimensions of the latter systems were not fixed, then the inputs in $\HS^{i_2}$ and $\HS^{i_3}$ could be transmitted via $B_1$ to $B_2$ which also receives the input from $\HS^{i_1}$. Then $B_2$ could trivially perform the unitary $U$ on the three input states and output one system to $\HS^{o_1}$ and the two others to $B_3$ which could eventually transmit them to $\HS^{o_2}$ and $\HS^{o_3}$. This would provide a decomposition of the unitary $U$ where the three gates $B_1,B_2$ and $B_3$ are unitaries and where the input and output systems are qubits. But then any tripartite unitary could be analogously ``decomposed'' in this setup, which highlights that fixing the dimension of the internal wires is crucial to study a non-trivial constraint of the circuit architecture on the decomposability of a given target unitary.

\subsection{SDP relaxation of Eq.~\eqref{eq:max_overlap_gen} in the main text}
\label{app:rewriting_relaxation_SDP}

We derive the relaxation presented in the main text of the optimisation problem in Eq.~\eqref{eq:max_overlap_gen} into the SDP problem presented in Eq.~\eqref{eq:max_overlap_cut_SDP}. As illustrated in Fig.~\ref{fig:U_2_gates}, consider the two gates $B_1:\HS^{i_1i_2}\to\HS^{o_1c}$ and $B_2:\HS^{c'i_3}\to\HS^{o_2o_3}$ composed according to $\Gamma=(\{1,2\},\{2,3\})$, where the systems in $\HS^c$ and $\HS^{c'}$ are connected by a qubit identity channel.  Mathematically, this internal wire is described by the Choi vector $\kket{\id}^{cc'}$ of the identity channel. Using the properties of the link product, we obtain the identity
\begin{align}
    \kketbra{B_1}{B_1}*\kketbra{B_2}{B_2} = (\kketbra{B_1}{B_1}\otimes\kketbra{B_2}{B_2})*\kketbra{\id}{\id}^{cc'},
\end{align}
which is illustrated in Fig.~\ref{fig:U_2_gates}b. Using this identity
one can rewrite Eq.~\eqref{eq:max_overlap_gen} in the main text as
    \begin{align}
         \textbf{given}\ \ & U, \Gamma \nonumber \\
    \textbf{max}_{B_1,B_2}\ \   & \frac{1}{64}\Tr\left[(\kketbra{U}{U}\otimes \kketbra{\id}{\id})\cdot (\kketbra{B_1}{B_1}\!\otimes\!\kketbra{B_2}{B_2}) \right] \nonumber \\
    \textbf{s.t.}\ \  & B_1,B_2 \ \text{are unitary operations,}
    \label{eq:max_overlap_cut}
    \end{align}
This rewriting allows to isolate the Choi matrices of the unitaries $B_1$ and $B_2$ while keeping track of the internal wire connecting these two gates. A relaxation on the unitaries can thus be performed without compromising the assumption that the internal wire should correspond to a qubit system. We thus perform a twofold relaxation of the problem by introducing a positive semidefinite variable $\tau  \in \L(\HS^{B_1^{IO}B_2^{IO}})$, and imposing positive partial transpose (PPT) on the bipartition $B_1^{IO}|B_2^{IO}$, as well as imposing that the marginals of $\tau$ on both sides of the bipartition are valid CPTP maps. The variable $\tau$ being positive semidefinite, the latter condition boils down to imposing only the trace preserving conditions. That way we retrieve the SDP problem defined in Eq.~\eqref{eq:max_overlap_cut_SDP} in the main text and reproduced here:
 \begin{align}
         \textbf{given}\ \ & U, \Gamma \nonumber \\
    \textbf{max}_\tau\ \   & \frac{1}{64}\Tr\left[(\kketbra{U}{U}\otimes \kketbra{\id}{\id})\cdot \tau\right] \nonumber \\
    \textbf{s.t.}\ \  & \tau \geq 0, \  \Tr[\tau] = 16, \nonumber \\
    &\tau^{T_{{B}^{IO}_1}}\geq0,\nonumber \\
    & \forall \, k=1,2, \ \Tr_{B_k^O}[\tau] = \Tr_{B_k^{IO}}[\tau]\otimes \frac{\id_{B_k^I}}{4}.
    \end{align}
We note that the relaxation could be made tighter in various ways. The PPT relaxation could be for instance replaced by the DPS hierarchy \cite{doherty03}, and the unitaries could be relaxed to be unital instead of CPTPs. Furthermore we highlight that the relaxed problem is of the form of the problems considered in \cite{berta22}, which provides a converging hierarchy of SDPs converging to the exact solution. We only used this formulation of the relaxation in the examples considered in this work as it was sufficient to witness the incompatibility we were looking for.

\subsection{SDP relaxation for the general case}
\label{app:rewriting_relaxation_SDP_general}

We now show that the SDP relaxation provided above for the minimal quantum circuit architecture $\Gamma=(\{1,2\},\{2,3\})$ can be extended to any quantum circuit architecture. Consider a target unitary operator $U:\HS^{i_1\ldots i_n}\to\HS^{o_1\ldots o_n}$ that transforms $n$ input systems respectively attached to input Hilbert spaces $\HS^{i_k}$ into $n$ output systems respectively attached to output Hilbert spaces $\HS^{o_k}$, for $k=1,\ldots,n$, and with the short-hand notation $\HS^{X_1\ldots X_n}:=\bigotimes_{k=1}^n\HS^{X_k}$. For the sake of simplicity we assume that all input and output Hilbert spaces have the same dimension $d$, i.e., $\HS^{i_k} \cong \HS^{o_k} \cong \mathbbm{C}^{d}$, for $k=1,\ldots,n$.
One would like to determine whether this target unitary $U$ can be decomposed as a sequential composition of $m$ unitary gates $B_1,\ldots,B_m$, see an illustration of the case $n=4$ and $m=5$ in Fig.~\ref{fig:causal_dec}. Importantly, we assume that each internal wire, which connects two unitary gates, is carrying a $d$-dimensional system. This assumption allows to label with integers (from 1 to $n$, top to bottom) the wires of the resulting circuit at any time step without ambiguity. A circuit architecture is then described by the list $\Gamma=(l_1,\ldots,l_m)$, where $l_k \subset \{1,\ldots,n\}$ denotes the labels of the systems on which is applied the gate $B_k:\HS^{B_k^I}\to\HS^{B_k^O}$ that transforms an incoming state in the input Hilbert space $\HS^{B_k^I} \cong \mathbbm{C}^{d \cdot|l_k|}$ into an outgoing state in the output Hilbert space $\HS^{B_k^O} \cong \mathbbm{C}^{d \cdot|l_k|}$. If such a decomposition exist, we say that the target unitary $U$ is compatible with the circuit architecture $\Gamma$.

Let $\kket{U} \in \HS^{i_1\ldots i_n o_1\ldots o_n}$ be the Choi vector of the target unitary $U$ with norm $d^n$. We want to optimise the fidelity between the Choi vector of the unitary transformation $\kket{U}$ and the Choi vector $\kket{B_1}*\ldots*\kket{B_m}$ describing any candidate quantum circuit compatible with the circuit architecture $\Gamma$. This boils down to the following fidelity optimisation problem:
    \begin{align}
         \textbf{given}\ \ & U, \Gamma \nonumber \\
    \textbf{max}_{B_1,\ldots,B_m}\ \   & \frac{1}{d^{2n}}\Tr\left[\kketbra{U}{U} \cdot \kketbra{B_1}{B_1}*\ldots*\kketbra{B_m}{B_m} \right] \nonumber \\
    \textbf{s.t.}\ \  & B_1,\ldots,B_m \ \text{are unitary operations applied according to the circuit architecture $\Gamma$.}
    \label{eq:max_overlap_gen_app}
    \end{align}

   We now rewrite the optimisation problem of Eq.~\eqref{eq:max_overlap_gen_app}, before relaxing it to a SDP problem. Given a circuit compatible with a circuit architecture, one crucial assumption is that all internal wires connecting two gates composing the circuit should represent a $d$-dimensional system. Analogously to the minimal example treated above, one way to make sure that even in a relaxation of the problem defined in Eq.~\eqref{eq:max_overlap_cut_SDP} this assumption is still satisfied is keep track of these internal wires explicitly. We thus ``open'' all the internal wires and explicitly connect them with an identity wire. More concretely, consider two gates $B_i$ and $B_j$, for $i<j$, that are connected according to the circuit architecture, i.e., $l_i \cap l_j \neq \{\}$. This implies that a physical system is wired from a subspace $\HS^x$ of $\HS^{B_i^O}$ to a subspace $\HS^{x'}$ of $\HS^{B_j^I}$. Mathematically, this wire is described by the Choi vector $\kket{\id}^{xx'}$ of the identity channel. Using the properties of the link product, we have the following identity
\begin{align}
    \kketbra{B_i}{B_i}*\kketbra{B_j}{B_j} = (\kketbra{B_i}{B_i}\otimes\kketbra{B_j}{B_j})*\kketbra{\id}{\id}^{xx'}.
\end{align}
This reformulation allows to isolate the Choi matrices of the unitaries $B_i$ and $B_j$ while keeping track of the internal wire(s) connecting these two gates. A relaxation on the unitaries can thus be performed without compromising the assumption that the internal wires should correspond to a $d$-dimensional system. Using this identity for each wire connecting two gates in the circuit architecture $\Gamma$, 
one can rewrite Eq.~\eqref{eq:max_overlap_gen_app} as
    \begin{align}
         \textbf{given}\ \ & U, \Gamma \nonumber \\
    \textbf{max}_{B_1,\ldots,B_m}\ \   & \frac{1}{d^{2n}}\Tr\left[(\kketbra{U}{U}\otimes \mathcal{I}_\Gamma)\cdot (\kketbra{B_1}{B_1}\!\otimes\!\ldots\!\otimes\!\kketbra{B_m}{B_m} )\right] \nonumber \\
    \textbf{s.t.}\ \  & B_1,\ldots,B_m \ \text{are unitary operations,}
    \label{eq:max_overlap_cut}
    \end{align}
with $\mathcal{I}_\Gamma=\kketbra{\id}{\id}\otimes\ldots\otimes \kketbra{\id}{\id}$, where each identity channel corresponds to an internal wire of the circuit architecture $\Gamma$. As such, the objective of Eq.~\eqref{eq:max_overlap_cut} is now proportional to the overlap between the (non-normalised) state $\kketbra{U}{U}\otimes \mathcal{I}_\Gamma$ and the (non-normalised) product state $\kketbra{B_1}{B_1}\otimes\ldots\otimes\kketbra{B_m}{B_m}$ according to the $m$-partition $B_1^{IO}|B_2^{IO}|\ldots |B_{m}^{IO}$, with the short-hand notation $B_k^{IO}:=B_k^{I}B_k^{O}$. It does not define an SDP problem as it is non-linear on the unitaries, and as imposing that the $B_k$ are unitaries is not a convex constraint. We thus relax the problem by introducing a positive semidefinite variable $\tau  \in \L(\HS^{B_1^{IO}\ldots B_m^{IO}})$, and imposing positive partial transpose on all bipartitions of the form $B_i^{IO}|B_1^{IO}\ldots B_{i-1}^{IO}B_{i+1}^{IO}\ldots B_{m}^{IO}$ for $i=1,\ldots,m$, as well as imposing that the gates $B_k$ are valid completely positive and trace-preserving (CPTP) maps. When $B_k$ are two-qudit gates, this defines the following SDP problem
    \begin{align}
         \textbf{given}\ \ & U, \Gamma \nonumber \\
    \textbf{max}_\tau\ \   & \frac{1}{d^{2n}}\Tr\left[(\kketbra{U}{U}\otimes \mathcal{I}_\Gamma)\cdot \tau\right] \nonumber \\
    \textbf{s.t.}\ \  & \tau \geq 0, \  \Tr[\tau] = d^{2m}, \nonumber \\
    & \forall \, k=1,\ldots,m, \ \tau^{T_{{B}^{IO}_k}}\geq0,\nonumber \\
    & \forall \, k=1,\ldots,m, \ \Tr_{B_k^O}\tau = \Tr_{B_k^{IO}}\tau\otimes \frac{\id_{B_k^I}}{d^2}.
\label{eq:SDP_gen_architecture}
    \end{align}


\section{An alternative relaxation for staircase circuit architectures}
\label{app:alternative_relaxation}

Building on the framework of so-called ``quantum combs''~\cite{Chiribella08,Chiribella09}, also known as ``quantum channels with memory''~\cite{Kretschmann05} or ``quantum strategies''~\cite{Gutoski06}, we propose an alternative relaxation to the optimisation problem presented in Eq.~\eqref{eq:max_overlap_gen} in the main text. We recall that this problem involves the certification that a target unitary $U$ is incompatible with the quantum circuit architecture $\Gamma=(\{1,2\},\{2,3\})$.

\subsection{Quantum comb framework}
\label{app:quantum_comb}
\begin{figure}[h]
\centering
\includegraphics[width=0.55\columnwidth]{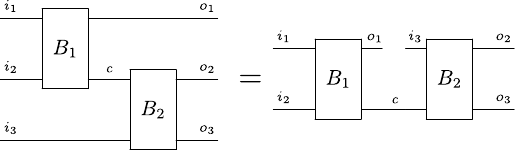}
    \caption{The quantum circuit defined as the composition of $B_1$ and $B_2$ according to the quantum circuit architecture $\Gamma=(\{1,2\},\{2,3\})$ can be rewritten in the shape of a quantum comb. Yet, as we imposed the dimension of the internal wire in $\HS^c$ to be a qubit, and as gates $B_1$ and $B_2$ are assumed to be unitaries, the right-hand side of the equation is a constrained quantum comb with a qubit memory system and unitary internal operations.}
    \label{fig:rewriting_comb}
\end{figure}

We consider the two gates $B_1:\HS^{i_1i_2}\to\HS^{o_1c}$ and $B_2:\HS^{ci_3}\to\HS^{o_2o_3}$ composed according to $\Gamma=(\{1,2\},\{2,3\})$, where the systems in $\HS^c$ is a qubit. As illustrated in Fig.~\ref{fig:rewriting_comb}, the circuit obtained by composing the two gates can be rewritten in a quantum comb shape. Leveraging the framework of quantum combs, and in particular the SDP characterisation that they admit, we are going to define an alternative way to witness the incompatibility of the target unitary with $\Gamma$. Importantly, we first note that dimension of the memory line in a quantum comb is arbitrarily large, while here we imposed that the Hilbert space $\HS^c$ corresponding to the internal wire is of dimension 2. Secondly, quantum combs are defined for internal operations $B_1$ and $B_2$ which are CPTPs and not necessarily unitaries. By optimising over quantum combs which have the shape of the one presented in Fig.~\ref{fig:rewriting_comb}, we are thus going to perform a two-fold relaxation of the problem presented in Eq.~\eqref{eq:max_overlap_gen} in the main text (arbitrary dimension of the internal wire and CPTP gates). The impossibility of decomposing a unitary as a quantum comb implying in particular the impossibility of decomposing the unitary as a constrained quantum comb (with dimension 2 memory and unitary gates), we will derive a witness out of the relaxation problem.

Interestingly, the Choi matrix of a quantum comb $W_{\text{comb}} \in \L(\HS^{i_1 i_2 i_3 o_1 o_2 o_3})$ admits the following characterisation, in terms of a positivity constraint and so-called trace-preserving constraints:
\begin{align}
    W_{\text{comb}} &\geq 0,\nonumber\\
    \Tr_{o_2o_3}W_{\text{comb}} &= \Tr_{i_3 o_2o_3}W_{\text{comb}}\otimes \frac{\id_{i_3}}{2},\nonumber\\
    \Tr_{o_1 o_2 o_3}W_{\text{comb}} &= \id_{i_1 i_2 i_3}.
    \label{eq:comb_charac}
\end{align}
We are now going to take advantage of the fact that this characterisation relies on linear constraints on a positive semidefinite matrix to relax the optimisation problem of Eq.~\eqref{eq:max_overlap_gen} into SDP problem.

\subsection{Building a witness}

We consider the architecture incompatibility problem presented in Eq.~\eqref{eq:max_overlap_gen} in the main text. An upper bound on the optimal fidelity can thus be obtained by optimising over the Choi matrix of the quantum comb characterised in Eq.~\eqref{eq:comb_charac} and solving the following problem 
\begin{align}
     \textbf{given}\ \ & U, \Gamma \nonumber \\
    \textbf{max}_{W_{\text{comb}}}\ \   & \frac{1}{64}\Tr\left[\kketbra{U}{U} \cdot W_{\text{comb}} \right] \nonumber \\
    \textbf{s.t.}\ \  & W_{\text{comb}} \ \text{satisfies Eq.~\eqref{eq:comb_charac},}
    \label{eq:max_overlap_comb}
\end{align}
which is a SDP problem and thus admits efficient numerical methods to be computed. We highlight that this SDP can be generalised to more general quantum circuit architectures than the one presented in Eq.~\eqref{eq:max_overlap_gen} in the main text, but does not straightforwardly generalise to all quantum circuit architectures contrarily to the approach presented in the main text. Indeed, by construction quantum combs assume no-signalling relations between the input and output systems (such as no-signalling from system $i_3$ to system $o_1$ for the example presented in Fig.~\ref{fig:rewriting_comb}), while general quantum circuit architectures allow for signalling from all input systems to all output systems. This explains why we can compare the approach presented here with the one presented in the main text for the examples of Table~\ref{tab:cascade_example_2_gates} (the architecture $\Gamma=(\{1,2\},\{2,3\})$ does not allow signalling from system $i_3$ to system $o_1$), but not for the examples of Table~\ref{tab:example_3_gates} (the architecture $\Gamma=(\{1,2\},\{2,3\},\{1,2\})$ allows for signalling from all input systems to all output systems).

As it turns out, we retrieve with this SDP the exact same upper bounds (up to numerical error) as the ones obtained with Eq.~\eqref{eq:max_overlap_cut_SDP} and presented in Table~\ref{tab:cascade_example_2_gates} in the main text. We point out that the agreement between the upper bounds of the two approaches is a priori not obvious as the two relaxations from Eq.~\eqref{eq:max_overlap_gen} to Eq.~\eqref{eq:max_overlap_cut_SDP} and to Eq.~\eqref{eq:max_overlap_comb} respectively are of different nature by construction. Both relaxations imply that the unitary gates are relaxed to CPTPs. Yet, the approach presented in the main text assumes a fixed dimension of the internal wire but it relaxes the product state structure of the variable $\kketbra{B_1}{B_1}\otimes \kketbra{B_2}{B_2}$ with the PPT constraint, while the approach presented here exclusively relaxes the dimension of the internal wire to be arbitrary. Identifying examples where the two relaxations lead to different upper bounds is an interesting question that we leave for future work.

Finally, we note that analogously to the rewriting from Eq.~\eqref{eq:max_overlap_cut_SDP} in the main text to Eq.~\eqref{eq:max_overlap_cut_LP}, the SDP defined in Eq.~\eqref{eq:max_overlap_comb} can also be turned into a LP when the target unitary is a Clifford one. The proof can be straightforwardly obtained from the one given in Appendix~\ref{app:rewriting_SDP_to_LP} below.

\section{Turning the SDP optimisation into a LP optimisation for Clifford unitaries}
\label{app:SDP_to_LP}

\subsection{Graph States and the Stabiliser Formalism}
\label{app:graph_stabiliser}

We review the basic features related to graph states and the stabiliser formalism that we employ in this paper, see Ref.~\cite{hein06} for a review. Let $G = (V,E)$ be a simple undirected graph with $|V|=n$. To each vertex $v \in V$ we associate a qubit. The \emph{graph state} $\ket{G}$ is defined as
\begin{equation}
\ket{G} = \prod_{(u,v)\in E} \mathrm{CZ}_{uv} \; \ket{+}^{\otimes n},
\end{equation}
where $\ket{+} = (\ket{0}+\ket{1})/\sqrt{2}$ and $\mathrm{CZ}_{uv}$ is the controlled-$Z$ gate acting on qubits $u$ and $v$. Equivalently, $\ket{G}$ is the unique simultaneous $+1$ eigenstate of the commuting set of operators
\begin{equation}
K_v = X_v \prod_{u \in N(v)} Z_u, \qquad v \in V,
\end{equation}
where $N(v)$ denotes the neighbourhood of $v$, and $X_v$, $Z_v$ are Pauli operators acting on qubit $v$. The operators $\{K_v\}_{v\in V}$ are called the \emph{stabilisers} of the graph.
The \emph{stabiliser group} of $\ket{G}$ is the Abelian group
\begin{equation}
\mathcal{S}(G) = \langle K_v \;:\; v\in V \rangle,
\end{equation}
consisting of all $2^n$ products of the generators $K_v$. The state $\ket{G}$ is uniquely characterised by
\begin{equation}
S \ket{G} = \ket{G}, \qquad \forall S \in \mathcal{S}(G).
\end{equation}

The generators $\{K_v\}$ can be used to define an orthonormal \emph{graph-state basis}. For any binary string $\mathbf{a}=(a_1,\dots,a_n)\in\{0,1\}^n$, define
\begin{equation}
\ket{G_{\mathbf{a}}} = \prod_{v\in V} Z_v^{a_v} \ket{G}.
\end{equation}
These states satisfy
\begin{equation}
K_v \ket{G_{\mathbf{a}}} = (-1)^{a_v} \ket{G_{\mathbf{a}}},
\end{equation}
and form a complete orthonormal basis of $(\mathbb{C}^2)^{\otimes n}$.

More generally, let $\mathcal{P}_n$ denote the $n$-qubit Pauli group. A pure $n$-qubit state $\ket{\psi}$ is called a \emph{stabiliser state} if there exists an Abelian subgroup $\mathcal{S} \subset \mathcal{P}_n$ such that:
(i) $-I \notin \mathcal{S}$,
(ii) $\mathcal{S}$ has $n$ independent generators,
and (iii) $\ket{\psi}$ is the unique simultaneous $+1$ eigenstate of all $S \in \mathcal{S}$:
\begin{equation}
S \ket{\psi} = \ket{\psi}, \qquad \forall S \in \mathcal{S}.
\end{equation}
Such a subgroup necessarily contains $2^n$ elements. Stabiliser states form a discrete subset of Hilbert space that is closed under Clifford unitaries. In particular this implies that by construction the Choi state of any Clifford unitary is a stabiliser state, as Bell states are stabiliser states.

Graph states constitute a particularly convenient subclass of stabiliser states. In fact, every graph state is a stabiliser state by construction. Conversely, any stabiliser state is \emph{local Clifford} (LC) equivalent to a graph state: for every stabiliser state $\ket{\psi}$, there exists a local Clifford unitary $U = \bigotimes_{v\in V} U_v$ such that
\begin{equation}
\ket{\psi} = U \ket{G}
\end{equation}
for some graph $G$. Thus, up to local Clifford transformations, graph states provide a normal form for all stabiliser states.

This correspondence allows us to translate problems about general stabiliser states into graph-theoretic statements about the associated graph, which we use as a technical tool in the following as well as in Appendix~\ref{app:analytical_witness}.

\subsection{Turning Eq.~\eqref{eq:max_overlap_cut_SDP} in the main text and Eq.~\eqref{eq:SDP_gen_architecture} into LPs}
\label{app:rewriting_SDP_to_LP}

We show that the SDP problem defined in Eq.~\eqref{eq:SDP_gen_architecture} to witness the incompatibility with a general quantum circuit architecture can be turned to linear programming if $U$ is a Clifford unitary applied on qubits ($d=2$). That Eq.~\eqref{eq:max_overlap_cut_SDP} in the main text can also be rewritten into a LP then follows automatically.

Assuming that $U$ is a Clifford unitary applied on qubits, we know that $\kket{U}$ is a stabiliser state, and so is $\kketbra{U}{U}\otimes \mathcal{I}_\Gamma$, see Appendix~\ref{app:graph_stabiliser}. Any stabiliser state is LC equivalent to a graph state, so there exists a local Clifford unitary $U_{\text{LC}}$ such that $\kketbra{U}{U}\otimes \mathcal{I}_\Gamma= U_{\text{LC}}\ketbra{G}{G}U_{\text{LC}}^\dagger$, $\ket{G}$ being a $p$-qubit graph state, where $p$ is a function of $n$ and $m$ which can be easily computed for a given quantum circuit architecture. Inserting this relation in Eq.~\eqref{eq:SDP_gen_architecture}, we obtain
 \begin{align}
         \textbf{given}\ \ & U,\Gamma \nonumber \\
    \textbf{max}_\tau\ \   & \frac{1}{2^{2n}}\Tr\left[\ketbra{G}{G}\cdot U_{\text{LC}}^\dagger\tau U_{\text{LC}}\right] \nonumber \\
    \textbf{s.t.}\ \  & U_{\text{LC}}^\dagger \tau U_{\text{LC}} \geq 0, \ \Tr[U_{\text{LC}}^\dagger \tau U_{\text{LC}}]=2^{2m}, \nonumber\\
    & \forall \, k=1,\ldots,m, \  (U_{\text{LC}}^\dagger \tau U_{\text{LC}})^{T_{B_k^{IO}}}\geq0, \notag \\
    & \forall \, k=1,\ldots,m, \   \Tr_{B_k^O}[U_{\text{LC}}^\dagger \tau U_{\text{LC}}] = \Tr_{B_k^{IO}}[U_{\text{LC}}^\dagger \tau U_{\text{LC}}] \otimes \frac{\id_{B_k^I}}{4}.
    \label{eq:app_max_overlap_SDP_2}
    \end{align}
Then, following the depolarisation procedure of Ref.~\cite{Jungnitsch11}, one can restrict the optimisation to operators that are diagonal in the graph-state basis. The procedure removes the off-diagonal components while preserving the overlap with the graph state and the relevant positivity, normalisation, PPT and trace-preserving constraints. Hence, Eq.~\eqref{eq:app_max_overlap_SDP_2} is maximised when $U_{\text{LC}}^\dagger \tau U_{\text{LC}}$ is diagonal in the graph-state basis defined by $\ket{G}$ (see Appendix~\ref{app:graph_stabiliser}). We can thus take without loss of generality $U_{\text{LC}}^\dagger \tau U_{\text{LC}} = \text{diag}(U_{\text{LC}}^\dagger \tau U_{\text{LC}})$, where $\text{diag}(\cdot)$ is the diagonal part of a given matrix. Eq.~\eqref{eq:app_max_overlap_SDP_2} is then equivalently rewritten as 
 \begin{align}
         \textbf{given}\ \ & U,\Gamma \nonumber \\
    \textbf{max}_\tau\ \   & \frac{1}{2^{2n}}\Tr\left[\ketbra{G}{G}\cdot \text{diag}(U_{\text{LC}}^\dagger\tau U_{\text{LC}})\right] \nonumber \\
    \textbf{s.t.}\ \  & \text{diag}(U_{\text{LC}}^\dagger\tau U_{\text{LC}}) \geq 0, \ \Tr[\text{diag}(U_{\text{LC}}^\dagger\tau U_{\text{LC}})]= 2^{2m} \nonumber\\
    & \forall \, k=1,\ldots,m, \ (\text{diag}(U_{\text{LC}}^\dagger\tau U_{\text{LC}}))^{T_{B_k^{IO}}}\geq0,\notag \\
    & \forall \, k=1,\ldots,m, \ \Tr_{B_k^O}[\text{diag}(U_{\text{LC}}^\dagger\tau U_{\text{LC}})] = \Tr_{B_k^{IO}}[\text{diag}(U_{\text{LC}}^\dagger\tau U_{\text{LC}})] \otimes \frac{\id_{B_k^I}}{4},
    \label{eq:app_max_overlap_SDP_3}
    \end{align}
with $\Tr[\mathcal{I}_\Gamma]^2\text{diag}(U_{\text{LC}}^\dagger\tau U_{\text{LC}}) = \sum_{\mathbf{a}\in \{0,1\}^p} \lambda_{\mathbf{a}} \ketbra{G_{\mathbf{a}}}{G_{\mathbf{a}}} = \sum_{\mathbf{s}\in \{0,1\}^p} \alpha_{\mathbf{s}} S_{\mathbf{s}} $, where $\ket{G_{\mathbf{a}}}$ is a generic vector of the graph-diagonal basis, $S_{\mathbf{s}}$ is a generic stabiliser of the graph state $\ket{G}$, $\lambda_{\mathbf{a}}$ are non-negative coefficients and $\alpha_{\mathbf{s}}$ are real coefficients. We take the convention that $\ket{G_{(0,\ldots,0)}}:=\ket{G}$. The two sets of coefficients are connected by the following discrete Fourier transform relations:
\begin{align}
    \alpha_{\mathbf{s}} &= \frac{1}{2^p} \sum_{\mathbf{a}} (-1)^{\mathbf{a}\cdot \mathbf{s}} \lambda_{\mathbf{a}}, \label{eq:alpha_to_lambda}\\
    \lambda_{\mathbf{a}} &= \sum_{\mathbf{s}} (-1)^{\mathbf{a}\cdot \mathbf{s}} \alpha_{\mathbf{s}}.\label{eq:lambda_to_alpha}
\end{align}

The two first constraints of Eq.~\eqref{eq:app_max_overlap_SDP_3} can then easily be rewritten as constraints on the coefficients $\lambda_{\mathbf{a}}$, namely positivity and normalisation of these coefficients. Using Eq.~\eqref{eq:lambda_to_alpha}, one can translate the last two conditions as constraints on $\alpha_{\mathbf{s}}$ rather than $\lambda_{\mathbf{a}}$, which turns out to be more convenient to simplify them. 
\begin{itemize}
    \item For the first condition, the partial transpose on the generic Hilbert space $\HS^{X_i}$ acts as follows on the $\alpha_{\mathbf{s}}$: if the corresponding stabiliser $S_{\mathbf{s}}$ has an even number of Pauli $Y$ strings, then $\alpha_{\mathbf{s}}$ is conserved, otherwise if the number of Pauli $Y$ strings is odd, then $\alpha_{\mathbf{s}}$ is transformed into $-\alpha_{\mathbf{s}}$. Once each $\alpha_{\mathbf{s}}$ is updated with a potential minus sign that we will denote with the function $\varepsilon_{X_i}(S_\mathbf{s})$, one can retrieve the condition on the coefficients $\lambda_{\mathbf{a}}$ using Eq.~\eqref{eq:alpha_to_lambda}.
    
    \item For the second condition, using the decomposition $\Tr[\mathcal{I}_\Gamma]^2\text{diag}(U_{\text{LC}}^\dagger\tau U_{\text{LC}}) = \sum_{\mathbf{s}\in \{0,1\}^p} \alpha_{\mathbf{s}} S_{\mathbf{s}}$, one can split the condition for each stabiliser as they are all independent. Then, for a given $\mathbf{s}$ the condition is non-trivial only if the reduced Pauli string on Hilbert space $\HS^{Y_j^O}$ of the corresponding stabiliser $S_\mathbf{s}$ is $\id_{Y_j^O}$, which we write as $S_\mathbf{s}^{Y_j^O} = \id_{Y_j^O}$ for short. Then, the condition imposes that for all the remaining $\alpha_{\mathbf{s}}$, the corresponding $S_\mathbf{s}^{Y_j^I}$ is identity. If this is not the case, it imposes that $\alpha_{\mathbf{s}}=0$, which can be easily translated as a condition on the coefficients $\lambda_{\mathbf{a}}$ using Eq.~\eqref{eq:alpha_to_lambda}.
\end{itemize}
Overall Eq.~\eqref{eq:app_max_overlap_SDP_3} can be rewritten as the following linear programming problem:

 \begin{align}
         \textbf{given}\ \ & U, \Gamma \nonumber \\
    \textbf{max}_{\{\lambda_{\mathbf{a}}\}_\mathbf{a}}\ \   & \lambda_{(0,\ldots,0)} \nonumber \\
    \textbf{s.t.}\ \  & \forall \, \mathbf{a}, \, \lambda_{\mathbf{a}} \geq 0, \ \sum_{\mathbf{a}} \lambda_{\mathbf{a}} = 2^{2m-n}\Tr[\mathcal{I}_\Gamma], \nonumber\\
    &\forall \, k=1,\ldots,m, \ \sum_{\mathbf{a},\mathbf{s}} (-1)^{\mathbf{a}\cdot \mathbf{s}}\varepsilon_{B_k^{IO}}(S_\mathbf{s})\lambda_{\mathbf{a}} \geq 0,\nonumber \\
    &\forall \, k=1,\ldots,m, \ \forall \, \mathbf{s}, \  \text{s.t.} \ S_{\mathbf{s}}^{B_k^O} = \id_{B_k^O} \ \text{and} \ S_{\mathbf{s}}^{B_k^I} \neq \id_{B_k^I}, \sum_{\mathbf{a}} (-1)^{\mathbf{a}\cdot \mathbf{s}} \lambda_{\mathbf{a}} = 0,
    \label{eq:max_overlap_cut_LP_general}
    \end{align}
where $\mathbf{a},\mathbf{s} \in \{0,1\}^{p}$, $S_{\mathbf{s}}$ are the stabilisers corresponding to the graph state $\ket{G}$, $S_{\mathbf{s}}^{B_k^I}$ (resp.~$S_{\mathbf{s}}^{B_k^O}$) denote the Pauli strings extracted from $S_{\mathbf{s}}$ that correspond to $\HS^{B_k^I}$ (resp.~$\HS^{B_k^O}$), and $\varepsilon_{B_k^{IO}}(S_\mathbf{s})$ is $+1$ if the number of Pauli $Y$ matrices in $S_{\mathbf{s}}^{B_k^{IO}}$ is even, and $-1$ otherwise.

For the sake of completeness, we write here the LP obtained from Eq.~\eqref{eq:max_overlap_cut_SDP} and corresponding to the minimal example studied in the main text, see an illustration in Fig.~\ref{fig:U_2_gates}a.
        \begin{align}
         \textbf{given}\ \ & U, \Gamma \nonumber \\
    \textbf{max}_{\{\lambda_{\mathbf{a}}\}_\mathbf{a}}\ \   & \lambda_{(0,\ldots,0)} \nonumber \\
    \textbf{s.t.}\ \  & \forall \, \mathbf{a}, \, \lambda_{\mathbf{a}} \geq 0, \ \sum_{\mathbf{a}} \lambda_{\mathbf{a}} = 4, \nonumber\\
    & \sum_{\mathbf{a},\mathbf{s}} (-1)^{\mathbf{a}\cdot \mathbf{s}}\varepsilon_{B_1^{IO}}(S_\mathbf{s})\lambda_{\mathbf{a}} \geq 0,\nonumber \\
    &\forall \, k=1,2, \ \forall \, \mathbf{s}, \  \text{s.t.} \ S_{\mathbf{s}}^{B_k^O} = \id_{B_k^O} \ \text{and} \ S_{\mathbf{s}}^{B_k^I} \neq \id_{B_k^I}, \notag \\
    &\quad \quad \sum_{\mathbf{a}} (-1)^{\mathbf{a}\cdot \mathbf{s}} \lambda_{\mathbf{a}} = 0,
    \label{eq:max_overlap_cut_LP}
    \end{align}
where $\mathbf{a},\mathbf{s} \in \{0,1\}^{8}$, $S_{\mathbf{s}}$ are the stabilisers corresponding to the graph state $\ket{G}$, $S_{\mathbf{s}}^{B_k^I}$ (resp.~$S_{\mathbf{s}}^{B_k^O}$) denote the Pauli strings extracted from $S_{\mathbf{s}}$ that correspond to $\HS^{B_k^I}$ (resp.~$\HS^{B_k^O}$), and $\varepsilon_{B_k^{IO}}(S_\mathbf{s})$ is $+1$ if the number of Pauli $Y$ matrices in $S_{\mathbf{s}}^{B_k^{IO}}$ is even, and $-1$ otherwise.

\section{Defining analytical witnesses}
\label{app:analytical_witness}

In this section we propose an analytical alternative technique to define witnesses, which is available for any Clifford target unitary $U$. We start by showing how the approach works for a minimal example, before showing that it allows us to deal with some quantum circuit architectures made of an arbitrary number of gates.  

\subsection{A minimal example}
\label{app:analytical_witness_2}
\begin{figure}[h]
\centering
\includegraphics[width=0.3\columnwidth]{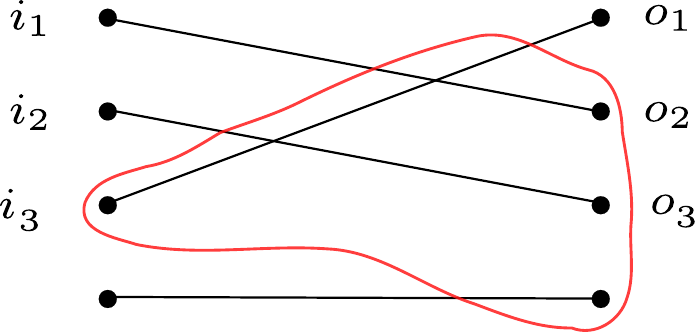}
\caption{Illustration of the graph state $\kketbra{U}{U}\!\otimes\!\kketbra{\id}{\id}$ when the target unitary $U$ is made of the composition of two SWAP gates arranged according to the staircase circuit architecture $\Gamma = (\{2,3\},\{1,2\})$. Each line corresponds to a Bell pair between the input and output systems. The bottom line corresponds to the identity wire added by cutting the circuit, which is described in the Choi picture by a Bell pair up to normalisation. The red line indicates the border of the bipartition $B_1^{IO}|B_{2}^{IO}$ considered in Eq.~\eqref{eq:app_ana_max_fide}.}
\label{fig:2_cascade_graph}
\end{figure}

We consider the following incompatibility problem: given a unitary $U$ made of two SWAP operations composed according to the architecture $\Gamma_{\text{target}} = (\{2,3\},\{1,2\})$, show the incompatibility of $U$ with the quantum circuit architecture $\Gamma_{\text{test}} = (\{1,2\},\{2,3\})$, see Fig.~\ref{fig:cascade_example_2_gates} for an illustration. We note that we already studied numerically this example and showed that the optimal fidelity is $\alpha_{\Gamma_{\text{test}}}(U) = 0.25$, see Table~\ref{tab:cascade_example_2_gates}. We now explain how we retrieve this value analytically. We reproduce the corresponding optimisation problem, without performing any relaxation:
  \begin{align}
         \textbf{given}\ \ & U, \Gamma \nonumber \\
    \textbf{max}_{B_1,B_2}\ \   & \frac{1}{64}\Tr\left[(\kketbra{U}{U}\otimes \kketbra{\id}{\id})\cdot (\kketbra{B_1}{B_1}\!\otimes\!\kketbra{B_2}{B_2}) \right] \nonumber \\
    \textbf{s.t.}\ \  & B_1,B_2 \ \text{are unitary operations applied according to the circuit architecture $\Gamma_{\text{test}}$.}
    \label{eq:}
    \end{align}
Contrarily to the relaxation performed above, we here replace $\kketbra{B_1}{B_1}\!\otimes\!\kketbra{B_2}{B_2}$ by a positive semidefinite variable $\tau_{1|2}$ which is constrained to be separable across the bipartition $B_1^{IO}|B_2^{IO}$:
 \begin{align}
         \textbf{given}\ \ & U, \Gamma \nonumber \\
    \textbf{max}_{\tau_{1|2}}\ \   & 4\Tr\left[\left(\frac{1}{16}\kketbra{U}{U}\otimes \kketbra{\id}{\id}\right)\cdot \tau_{1|2} \right] \nonumber \\
    \textbf{s.t.}\ \  & \tau_{1|2} \in \text{SEP}(B_1^{IO}|B_2^{IO})
    \label{eq:app_ana_max_fide}
    \end{align}
where $\text{SEP}(B_X^{IO}|B_Y^{IO}) \subset \L(\HS^{B_X^{IO}B_Y^{IO}})$ is the set of (normalised) quantum states which are separable across the bipartition $B_X^{IO}|B_Y^{IO}$.
From the requirement that all $B_k$ are unitaries, we only kept the positivity and normalisation of their Choi matrices in this relaxation, while we for instance did not keep the trace-preserving condition. We now explain how to compute analytically this maximisation by employing the stabiliser state formalism. In this example the stabiliser state $\kketbra{U}{U}\!\otimes\!\kketbra{\id}{\id}$ is already a graph state, and can thus be represented graphically, see Fig.~\ref{fig:2_cascade_graph}.
Since this graph state is a collection of Bell pairs, the optimal fidelity defined in Eq.~\eqref{eq:app_ana_max_fide} can be obtained simply by counting the number of Bell pairs that are defined across the bipartition $B_1^{IO}|B_{2}^{IO}$. Indeed, the overlap with each Bell pair defined across the bipartition is at most 1/2, since the maximal overlap between a Bell pair and a two-qubit separable state is 1/2. In that case 4 Bell pairs are defined across the bipartition (see Fig.~\ref{fig:2_cascade_graph}), each contributing with a factor 1/2. Taking into account the normalisation, we thus obtain the analytical upper bound 
\begin{align}
\alpha_{\Gamma_{\text{test}}}(U) \leq 4\times \left(\frac{1}{2}\right)^4 = \frac{1}{4}.
\end{align}
We thus retrieve the upper bound obtained numerically in Table~\ref{tab:cascade_example_2_gates}. We note that in the case where the Choi state $\kketbra{U}{U}\!\otimes\!\kketbra{\id}{\id}$ is a stabiliser state but not a graph state, then one can employ the reduction technique from Ref.~\cite{Fattal04} to transform the stabiliser state into a collection of Bell pairs and local terms, using local (in the sense of the bipartition) unitaries. As the fidelity with local terms can always be optimised to be 1, the argument about counting Bell pairs just explained also provides an analytical upper bound in that case.

In the next subsection, we propose a generalisation of this example, considering an analogous quantum circuit architecture made of an arbitrary number $m$ of gates.

\subsection{Generalisation to a $m$-gates architecture}
\label{app:analytical_witness_N}

\begin{figure}[h!]
\centering
\includegraphics[width=0.95\columnwidth]{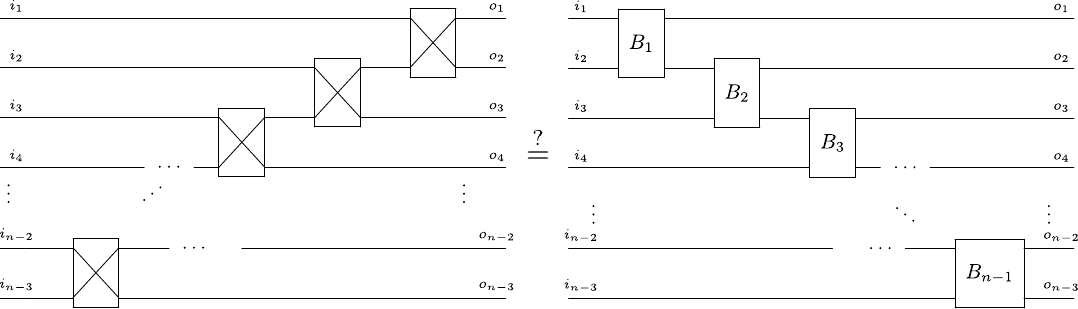}
    \caption{Graphical representation of a quantum circuit made of the composition of $m=n-1$ gates $B_k$ arranged according to the quantum circuit architecture $\Gamma = (\{1,2\},\{2,3\},\ldots,\{n-1,n\})$.}
    \label{fig:quantum_comb}
\end{figure}

To illustrate this technique, we consider the example of a target unitary $U$ made of the composition of $m=n-1$ SWAP gates arranged according to a staircase quantum circuit architecture $\Gamma_{\text{target}}^n = (\{n-1,n\},\{n-2,n-1\},\ldots,\{1,2\})$. The test circuit architecture is then chosen to be the converse staircase quantum architecture $\Gamma_{\text{test}}^n = (\{1,2\},\{2,3\},\ldots,\{n-1,n\})$ made of $m=n-1$ two-qubit gates applied on $n$ qubits, see an illustration in Fig.~\ref{fig:quantum_comb}. 

Instead of using the rewriting done in Eq.~\eqref{eq:max_overlap_cut}, which involved cutting all internal wires, we here only cut the internal wire between $B_{n-2}$ and $B_{n-1}$, according to
\begin{align}
    &\Tr\left[\kketbra{U}{U} \cdot \kketbra{B_1}{B_1}*\ldots*\kketbra{B_{n-1}}{B_{n-1}} \right] = \Tr\bigl[(\kketbra{U}{U}\!\otimes\! \kketbra{\id}{\id})\!\cdot (\kketbra{B_1}{B_1}\!*\!\ldots\!*\!\kketbra{B_{n-2}}{B_{n-2}}\!\otimes\!\kketbra{B_{n-1}}{B_{n-1}})\bigl].
\end{align}

We thus obtain the following optimisation problem:
    \begin{align}
         \textbf{given}\ \ & U, \Gamma \nonumber \\
    \textbf{max}_{B_1,\ldots,B_{n-1}}\ \   & \frac{1}{2^{2n}}\Tr\left[(\kketbra{U}{U}\otimes \kketbra{\id}{\id})\cdot (\kketbra{B_1}{B_1}\!*\!\ldots\!*\!\kketbra{B_{n-2}}{B_{n-2}}\!\otimes\!\kketbra{B_{n-1}}{B_{n-1}})\right] \nonumber \\
    \textbf{s.t.}\ \  & B_1,\ldots,B_{n-1} \ \text{are unitary operations.}
    \label{eq:}
    \end{align}

\begin{figure}[h]
\centering
\includegraphics[width=0.3\columnwidth]{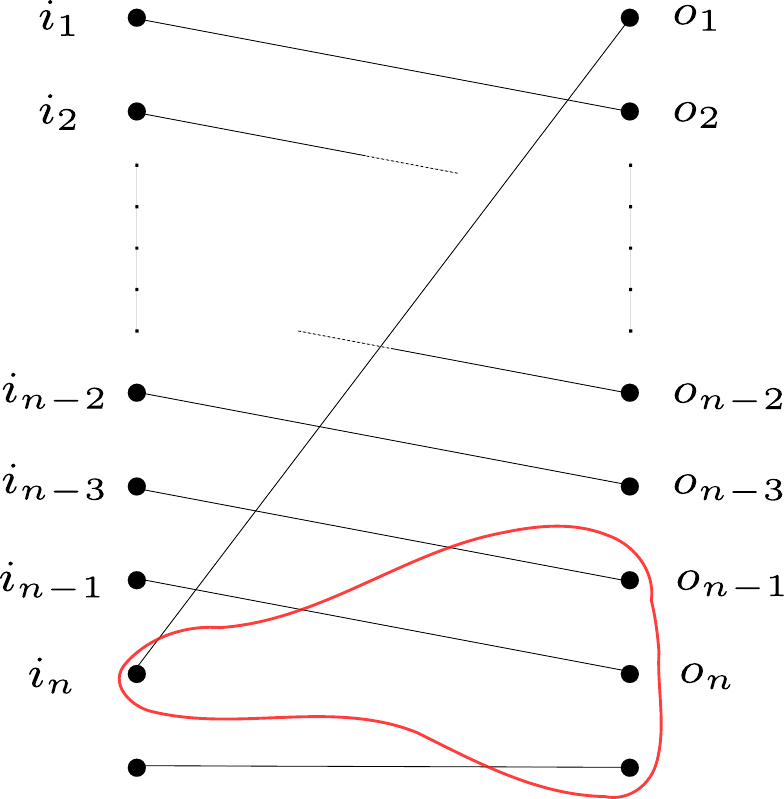}
\caption{Illustration of the graph state $\kketbra{U}{U}\!\otimes\!\kketbra{\id}{\id}$ when the target unitary $U$ is made of the composition of $n-1$ SWAP gates arranged according to the staircase circuit architecture $\Gamma = (\{n-1,n\},\{n-2,n-1\},\ldots,\{1,2\})$. Each line corresponds to a Bell pair between the input and output systems. The bottom line corresponds to the identity wire added by cutting the circuit, which is described in the Choi picture by a Bell pair up to normalisation. The red line indicates the border of the bipartition $B_1^{IO}\!\ldots B_{n-2}^{IO}|B_{n-1}^{IO}$ considered in Eq.~\eqref{eq:max_overlap_relax_cascade_analytical}.}
\label{fig:N_cascade_graph}
\end{figure}

As already observed with the rewriting of Eq.~\eqref{eq:max_overlap_cut}, the optimisation problem becomes a maximisation of the fidelity between the entangled state $\kketbra{U}{U}\otimes \kketbra{\id}{\id}$ and a product state $\kketbra{B_1}{B_1}\!*\!\ldots\!*\!\kketbra{B_{n-2}}{B_{n-2}}\!\otimes\!\kketbra{B_{n-1}}{B_{n-1}}$ across the $B_1^{IO}\!\ldots B_{n-2}^{IO}|B_{n-1}^{IO}$ bipartition, with the extra constraints that the $B_k$ are unitaries. As already stated above, since here $U$ is a Clifford unitary operation then $\kket{U}$ is a stabiliser state.
A relaxation on the optimisation problem is obtained by optimising over quantum states $\tau_{1\ldots n-2|n-1} \in \L(\HS^{B_1^{IO}\ldots B_{n-1}^{IO}})$ that are separable across the bipartition $B_1^{IO}\!\ldots B_{n-2}^{IO}|B_{n-1}^{IO}$:
  \begin{align}
         \textbf{given}\ \ & U, \Gamma \nonumber \\
    \textbf{max}_{\tau_{1\ldots n-2|n-1}}\ \   & 4\Tr\left[\left(\frac{1}{2}\kketbra{U}{U}\otimes \kketbra{\id}{\id}\right)\cdot\!\tau_{1\ldots n-2|n-1}\right] \nonumber \\
    \textbf{s.t.}\ \  &\tau_{1\ldots n-2|n-1} \in \text{SEP}(B_1^{IO}\!\ldots B_{n-2}^{IO}|B_{n-1}^{IO}),
\label{eq:max_overlap_relax_cascade_analytical}
    \end{align}
From the requirement that all $B_k$ are unitaries, we only kept the positivity and normalisation of their Choi matrices in this relaxation. In that example the stabiliser state $\kketbra{U}{U}\!\otimes\!\kketbra{\id}{\id}$ is already a graph state, and can thus be represented graphically, see Fig.~\ref{fig:N_cascade_graph}. 
Since this graph state is a collection of Bell pairs, the optimal fidelity can be obtained simply by counting the number of Bell pairs that are defined across the bipartition $B_1^{IO}\!\ldots B_{n-2}^{IO}|B_{n-1}^{IO}$. Indeed, the overlap with each Bell pair that is not defined across the bipartition can be easily maximised to 1, while the overlap with the Bell pairs defined across the bipartition is at most 1/2, since the maximal overlap between a Bell state and a two-qubit separable state is 1/2. In that case 4 Bell pairs are defined across the bipartition (see Fig.~\ref{fig:N_cascade_graph}), each contributing with a factor 1/2. Taking into account the normalisation, we thus obtain the analytical upper bound 
\begin{align}
\alpha_{\Gamma_{\text{test}}^n}(U) \leq 4\times \left(\frac{1}{2}\right)^4 = \frac{1}{4}, \ \text{for any} \ n.
\end{align}
We thus retrieve the upper bound obtained in Appendix~\ref{app:analytical_witness_2} for the case $n=3$, and show that this result actually generalises to any number $m=n-1$ of gates.

The same technique allows us to test the compatibility of any Clifford unitary $U$ with a staircase circuit architecture, and to bound analytically the optimised fidelity. An interesting follow-up question is whether the technique can also be considered to test the compatibility of any Clifford unitary with a general circuit architecture that does not assume any no-signalling relation between the input and output systems. Some preliminary results suggest that the technique needs some refinement to provide non-trivial analytical upper bounds in that case. In particular, in order to obtain a non-trivial upper bound one can no longer relax the Choi matrices of the unitaries to be only positive and normalised, but one should take into account some trace-preserving constraints in the relaxation, as was done in the main text in Eq.~\eqref{eq:max_overlap_cut_SDP}. These extra constraints imply that the upper bound is no longer directly given by a counting of the number of Bell pairs that can be extracted across a given bipartition. It remains work in progress to provide a systematic technique to upper bound analytically the optimised fidelity for a generic quantum circuit architecture.
\end{document}